\documentclass{aa}
\usepackage{txfonts}
\usepackage{graphicx}
\usepackage{natbib}
\bibpunct{(}{)}{;}{a}{}{,}

\begin{document}

\newcommand{\dd}{\mathrm{d}}
\newcommand{\eps}{\varepsilon}
\newcommand{\crp}{\mathrm{CRp}}
\newcommand{\cre}{\mathrm{CRe}}
\newcommand{\e}{\mathrm{e}}
\newcommand{\p}{\mathrm{p}}
\newcommand{\me}{m_\e}

\title{Probing the Cosmic Ray Population of the Giant Elliptical Galaxy M~87
  with Observed TeV $\gamma$-Rays}
\titlerunning{Probing the Cosmic Ray Population of the Giant Elliptical Galaxy M~87}
\author{Christoph Pfrommer \and Torsten A. En{\ss}lin}
\authorrunning{C. Pfrommer \& T. A. En{\ss}lin}
\institute{Max-Planck-Institut f\"{u}r Astrophysik,
Karl-Schwarzschild-Str.1, Postfach 1317, 85741 Garching, 
Germany} 
\offprints{Christoph Pfrommer, \\
\email{pfrommer@mpa-garching.mpg.de}}
\date{Submitted May 26, 2003 / Accepted July 15, 2003}

\abstract{We examine the cosmic ray proton (CRp) population within the giant
  elliptical galaxy M~87 using the TeV $\gamma$-ray detection of the HEGRA
  collaboration. In our scenario, the $\gamma$-rays are produced by decaying
  pions which result from hadronic CRp interactions with thermal gas of the
  interstellar medium of M~87. By comparing the $\gamma$-ray emission to upper
  limits from EGRET, we constrain the spectral index of the CRp population to
  $\alpha_\mathrm{GeV}^\mathrm{TeV}<2.275$ within our scenario. Both the
  expected radial $\gamma$-ray profile and the required amount of CRp
  support this hadronic scenario. The accompanying radio mini-halo of
  hadronically originating cosmic ray electrons is outshone by the synchrotron
  emission of the relativistic jet of M~87 by one order of magnitude. According
  to our predictions, the future GLAST mission should allow us to test this
  hadronic scenario.
}
\maketitle

\section{Introduction}
\label{sec:intro}

The giant elliptical galaxy \object{M~87} is an intensively studied object in
our direct extragalactic vicinity situated at a distance of 17~Mpc
\citep{2000ApJ...536..255N}. The announcement of the TeV $\gamma$-ray detection
of M~87 at a 4-$\sigma$ significance level by the HEGRA collaboration
\citep{2003A&A...403L...1A} using imaging atmospheric \v{C}erenkov techniques
was the first discovery of TeV $\gamma$-rays from a radio galaxy with a jet
whose axis forms a relatively large angle with the line of sight of roughly
$30\degr - 35\degr$ \citep{1996ApJ...467..597B}. On the basis of the limited
event statistics the detected emission is inconclusive whether it originates
from a point source or an extended source. Despite testing for burstlike
behavior of M~87 no time variation of the TeV $\gamma$-ray flux has been found.
This detection provides the unique possibility for probing different
$\gamma$-ray emission scenarios and thus provides new astrophysical insight
into high energy phenomena of this class of objects.

In the literature, there are three different types of model predicting
$\gamma$-ray emission from objects like M~87: In the first scenario, the
GeV/TeV $\gamma$-ray emission is generated by the active galactic nucleus
(AGN), and possibly related to processed radiation of the relativistic outflow
\citep{1997ApJS..109..103D}. Particularly, inverse Compton (IC) scattering of
cosmic microwave background photons off electrons within the jet which have
been directly accelerated or reaccelerated as well as the Synchrotron Self
Compton scenario could lead to $\gamma$-ray emission
\citep{2001ApJ...549L.173B}.  Secondly, dark matter annihilation or decay
processes could be another conceivable source of $\gamma$-ray emission, such as
the hypothetical neutralino annihilation \citep{2000PhRvD..61b3514B}.  Finally,
hadronic cosmic ray proton (CRp) interactions with the thermal ambient gas
would produce pion decay induced $\gamma$-rays as well as inverse Compton and
synchrotron emission by secondary cosmic ray electrons (CRe)
\citep{1982AJ.....87.1266V}.  These processes are possible due to the long
lifetimes of CRp comparable to the Hubble time \citep{1996SSRv...75..279V},
long enough to diffuse away from the production site and to maintain their
distribution throughout the cluster volume.  Because of the strong dependence
of this hadronic process on particle density, the giant radio galaxy M~87,
located inside the central cooling flow region of the \object{Virgo} cluster,
is expected to be a major site of $\gamma$-ray emission \citep{2003A&A...P}.

This work uses the hadronic scenario to model the resulting $\gamma$-ray
emission. Thus it probes the CRp population by the recent TeV $\gamma$-ray
observations yielding either an upper limit or a detection on the CRp
population, provided this scenario applies. However, this approach only
constrains the CRp within the central region of intracluster medium (ICM) of
the Virgo cluster which is dominated by the interstellar medium (ISM) of the
radio galaxy M~87. In the following, we use the term ICM for both. It should be
emphasized that this hadronic scenario predicts stationary $\gamma$-ray
emission and will be ruled out if the emission is found to be time-variable
(barring the existence of a second component). This, however, would result in
even tighter constraints on the CRp population owing to the absence of
inescapably accompanying $\gamma$-ray emission.

\section{$\gamma$-ray emission from hadronic CRp interactions}
\label{sec:theory}

The differential number density distribution of a CRp population can be
described by a power-law in momentum $p_\p$,
\begin{equation}
\label{fp}
f_\p (\vec{r}, p_\p) \,\dd p_\p\,\dd V = 
\tilde{n}_\crp(\vec{r})\, \left(\frac{p_\p \,c}{\mbox{GeV}} \right)^{-\alpha}\,
\left(\frac{c\,\dd p_\p}{\mbox{GeV}}\right)\, \dd V\,,
\end{equation}
where the normalization $\tilde{n}_\crp(\vec{r})$ is determined by different
models of spatial distribution of the CRp population (see
Sect.~\ref{CRpdistribution}).

If the CRp population within the cooling flow region had time to loose energy
by means of Coulomb interactions in the plasma \citep{1972Physica....58..379G},
the low energy part of the spectrum would be modified. This can be treated
approximately by imposing a lower momentum cutoff
\begin{equation}
\label{pmin}
p_\mathrm{min} = \beta_\p \gamma_\p \,m_\p\,c \simeq
2.0\,\left(\frac{t_\mathrm{age}}{\mbox{Gyr}} \right)
\left(\frac{n_\e}{0.1\,\mbox{cm}^{-3}} \right)\, \mbox{GeV} c^{-1},
\end{equation}
where we inserted typical values for M~87.  The kinetic energy density of such
a CRp population is
\begin{eqnarray}
\label{energydensity}
\eps_\crp(\vec{r}) &=& 
\frac{\tilde{n}_\crp(\vec{r})\,m_\p \,c^2}{2\,(\alpha-1)}\,
\left(\frac{m_\p \,c^2}{\mbox{GeV}}\right)^{1-\alpha}\times \\
&&\left[\mathcal{B}_{x}\left(\frac{\alpha-2}{2},
\frac{3-\alpha}{2}\right) + 2\,\tilde{p}^{1-\alpha} 
\left( \sqrt{1+\tilde{p}^2} - 1 \right)\right],\nonumber
\end{eqnarray}
where $\mathcal{B}_x(a,b)$ denotes the incomplete beta-function,  $x =
(1+\tilde{p}^2)^{-1}$, and $\tilde{p} = \frac{p_\mathrm{min}}{m_\p\,c}$.

The CRp interact hadronically with the thermal ambient gas and produce pions,
provided their momentum exceeds the kinematic threshold $p_\mathrm{thr} = 0.78
\mbox{ GeV }c^{-1}$ of the reaction.  The neutral pions decay into
$\gamma$-rays while the charged pions decay into secondary electrons (and
neutrinos).  Only the CRp population above the kinematic threshold
$p_\mathrm{thr}$ is visible through its decay products in $\gamma$-rays and
thus constrained by this work while its lower energy part can not in general be
limited by considering hadronic interactions only.

An analytic formula describing the omnidirectional (i.e. integrated over
$4\,\pi$ solid angle) differential $\gamma$-ray source function resulting from
$\pi^0$-decay is given in \citet{2003A&A...P}:
\begin{eqnarray}
\label{q gamma}
\lefteqn{
q_\gamma(\vec{r},E_\gamma)\,\dd E_\gamma\,\dd V\simeq
\sigma_\mathrm{pp}\,c\,n_\mathrm{N}(\vec{r})\,
2^{2-\alpha}\,\frac{\tilde{n}_\crp(\vec{r})}{\mbox{GeV}}\,\times}  \\
  & & \frac{4}{3\,\alpha}\,\left( \frac{m_{\pi^0}\,c^2}
      {\mbox{GeV}}\right)^{-\alpha}
      \left[\left(\frac{2\, E_\gamma}{m_{\pi^0}\, c^2}\right)^{\delta} +
      \left(\frac{2\, E_\gamma}{m_{\pi^0}\, c^2}\right)^{-\delta}
      \right]^{-\alpha/\delta}\!\dd E_\gamma\,\dd V\,, \nonumber
\end{eqnarray}
where $n_\mathrm{N}(\vec{r})$ is the target nucleon density in the ICM assuming
primordial element composition.  The formalism also includes the detailed
physical processes at the threshold of pion production like the velocity
distribution of CRp, momentum dependent inelastic CRp-p cross section, and kaon
decay channels.  The shape parameter $\delta$ and the effective cross section
$\sigma_\mathrm{pp}$ depend on the spectral index of the $\gamma$-ray spectrum
according to
\begin{eqnarray}
\label{delta}
\delta &=& 0.14 \,\alpha^{-1.6} + 0.44\qquad\mbox{and} \\
\label{sigmapp}
\sigma_\mathrm{pp} &=& 32 \cdot
\left(0.96 + \mathrm{e}^{4.4 \,-\, 2.4\,\alpha}\right)\mbox{ mbarn}\,. 
\end{eqnarray}

Provided the CRp population has a power-law spectrum, the relation of the
hadronic $\gamma$-ray flux $\mathcal{F}_\gamma$ in different energy bands can
easily be found using the analytic formulae for the integrated $\gamma$-ray
source density \citep{2003A&A...P},
\begin{eqnarray}
\frac{\mathcal{F}_\gamma (E_1 < E_\gamma < E_2)}{
\mathcal{F}_\gamma (E_3 < E_\gamma < E_4)}
&=& \frac{A_\gamma(E_1, E_2)}{A_\gamma(E_3, E_4)}\,, \\
\mbox{where~~} A_\gamma(E_i,E_j) &=&
\left[\mathcal{B}_x\left(\frac{\alpha + 1}{2\,\delta},
      \frac{\alpha - 1}{2\,\delta}\right)\right]_{x_i}^{x_j} 
\label{Ngamma}\\
\mbox{and~~}
x_{i,j} &=& \left[1+\left(\frac{m_{\pi^0}\,c^2}{2\,E_{i,j}}
      \right)^{2\,\delta}\right]^{-1}.
\end{eqnarray}
Here we introduced the abbreviation $[f(x)]_{x_1}^{x_2} = f(x_2) - f(x_1)$.
This relation for hadronic $\gamma$-ray fluxes is independent of any specific
model of CRp spatial distribution as long as the same physical mechanism
governs the CRp distribution in both energy bands.

Using the HEGRA $\gamma$-ray flux for M 87 of $\mathcal{F}_\gamma (E > 730
\mbox{ GeV}) = 9.6 \times 10^{-13}\, \gamma\mbox{ cm}^{-2} \mbox{ s}^{-1}$
\citep{2003A&A...403L...1A}, and requiring the expected $\gamma$-ray flux above
100~MeV to be smaller than the EGRET upper limit $\mathcal{F}_\gamma (E > 100
\mbox{ MeV}) = 2.18 \times 10^{-8}\, \gamma\mbox{ cm}^{-2} \mbox{ s}^{-1}$
\citep{2003ApJ...588..155R}, we are able to constrain the CRp spectral index to
$\alpha~<~2.275$.  For this calculation, we assume a constant CRp spectral
index $\alpha_\mathrm{GeV}^\mathrm{TeV}$ extending from the GeV to TeV energy
regime.  In the case of steeper spectra in the TeV region, the CRp spectrum
needs to be bent in a convex fashion or to exhibit a low energy cutoff in order
to meet the requirement imposed by EGRET.

\section{Spatial distributions of CRp within the ICM}
\label{CRpdistribution}
In the following, we introduce three models for the spatial distribution of CRp
within the ICM. The origin of the CRp population is not specified in the first
two models, but the CRp may be accelerated by shock waves of cluster mergers,
accretion shocks, or result from supernova driven galactic winds.

The {\bf isobaric model} assumes that the average kinetic CRp energy density
$\eps_{\crp} (\vec{r})$ is a constant fraction of the thermal energy density
$\eps_\mathrm{th} (\vec{r})$ of the ICM
\begin{equation}
\eps_\crp(\vec{r}) =  X_\crp \, \eps_\mathrm{th}(\vec{r})\,.
\end{equation}
The thermal energy density $\eps_\mathrm{th}(\vec{r})$ is obtained from a
spherically symmetric temperature profile \citep{2003A&A...P} and a double
$\beta$-model of electron densities adapted to X-ray observations of M~87
\citep{2002A&A...386...77M}.

The {\bf adiabatic model} assumes the CRp population to be originally isobaric
to the thermal population but to become adiabatically compressed during the
formation of the cooling flow without relaxing afterwards:
\begin{equation}
\eps_\crp(\vec{r}) = X_\crp\,\eps_\mathrm{th}(\vec{r}) \;\to\;
\eps'_\crp(\vec{r}') = X'_\crp(\vec{r}')\,\eps_\mathrm{th}(\vec{r}')\,.
\end{equation}
Adiabatic compression of CRp changes the scaling parameter $X_\crp$ to
\begin{equation}
\label{Xcrp adiabatic}
X'_\crp(\vec{r}) = 
X_\crp \,\left(\frac{n'_\e(\vec{r})}{n_\e(\vec{r})} \right)^{(\alpha + 2)/3} = 
X_\crp \,\left(\frac{T_\mathrm{cluster}}
{T_\e'(\vec{r})} \right)^{(\alpha + 2)/3} \,, 
\end{equation}
where $T_\mathrm{cluster}$ denotes the electron temperature in the outer region
of Virgo. The last step assumes hydrostatic equilibrium of the gas during this
transition.

{\bf Diffusion of CRp away from M~87:} The relativistic plasma bubbles produced
by M~87 likely contain relativistic protons, which can partly escape into the
thermal ICM \citep{2003A&A...399..409E}.  Most of the CRp that have been
injected into the cluster center are either diffusively transported into the
surrounding ICM or form relativistic bubbles which rise in the gravitational
potential of the cluster due to buoyant forces \citep[][ and references
therein]{2001ApJ...554..261C}.

Momentum dependent CRp diffusion in a turbulent magnetic field with a
Kolmogorov-type spectrum on small scales would result in spectral steepening
and therefore would violate the limits on the spectral index
$\alpha_\mathrm{GeV}^\mathrm{TeV}$ obtained in Sect.~\ref{sec:theory} provided
there is no sharp upper cutoff in the CRp spectrum. Thus we adopt for
simplicity the scenario of passive advective transport of CRp in a turbulent
flow with a diffusion coefficient $\kappa$ independent of momentum.  The
time-dependent CRp distribution function reads for short (duration $\Delta t$)
point-like injection with CRp injection rate $Q(p_\p)$ at time $t=0$
\begin{equation}
  \label{eq:time-dependent}
  f_\p(r,p_\p,t) = \frac{Q(p_\p)\, \Delta t}{(4 \pi\, t\, \kappa)^{3/2}}\, 
                   \exp\left(-\frac{r^2}{4\, t\, \kappa}\right)\,.
\end{equation}
In a quasi-stationary situation, which is a valid approximation for timescales
longer than the typical CRp diffusion timescale in the case of a stationary or
short-term intermittent CRp source, the integrated CRp distribution function
is approximately given by
\begin{equation}
  \label{eq:fp_integrated}
  f_\p(r,p_\p) = \int_0^t \!\dd t' f_\p(r,p_\p,t') = 
    \frac{Q(p_\p)}{4 \pi\, r\, \kappa}\,
    \mathrm{erfc}\left(\frac{r}{\sqrt{4\, t\, \kappa}}\right)\,,    
\end{equation}
where $\mathrm{erfc}(x)$ denotes the complementary error function which is
responsible for the spatial cutoff at the characteristic diffusion scale
$R_\mathrm{diff} = \sqrt{6\, t\,\kappa}$. While assuming a power-law shaped
$Q(p_\p)$, the CRp distribution of eq.~(\ref{fp}) can be written within the
framework set by this model as
\begin{equation}
  \label{eq:fp_diff}
  f_\p(r,p_\p) = 
    \frac{\tilde{n}_{\crp,0}\, c}{\mbox{GeV}}\,
    \left(\frac{r}{\mbox{kpc}}\right)^{-1}
    \mathrm{erfc}\left(\frac{r}{\sqrt{4\, t\, \kappa}}\right)
    \left(\frac{p_\p\,c}{\mbox{GeV}}\right)^{-\alpha}.    
\end{equation}

Following \citet{2003A&A...P}, the averaged CRp luminosity of M~87 can be
estimated to be
\begin{eqnarray}
\label{LCRp}
\frac{L_\crp}{\kappa} = \frac{4\pi\,m_\p c^2\,
\tilde{n}_{\crp,0}\mbox{ kpc}}{2\,(\alpha-1)}\,
\left(\frac{m_\p c^2}{\mbox{GeV}} \right)^{1-\alpha}\!
\mathcal{B}\left(\frac{\alpha-2}{2},
\frac{3-\alpha}{2}\right).
\end{eqnarray}
Considering an energy dependent mean-free-path for diffusion would require
fine-tuning of this model while simultaneously enlarging the accessible
parameter space. While the resulting profiles should only be slightly affected
by this change, this could possibly alter our conclusions concerning the
normalization $\tilde{n}_{\crp,0}$. However, this would only add further
uncertainty to $\tilde{n}_{\crp,0}$ which already depends on two unknown
parameters, the lifetime of the source $t$ and the diffusion coefficient
$\kappa$.

\section{Modeled $\gamma$-ray profiles}
\label{gamma profiles}

\begin{figure}[t]
\resizebox{\hsize}{!}{
\includegraphics{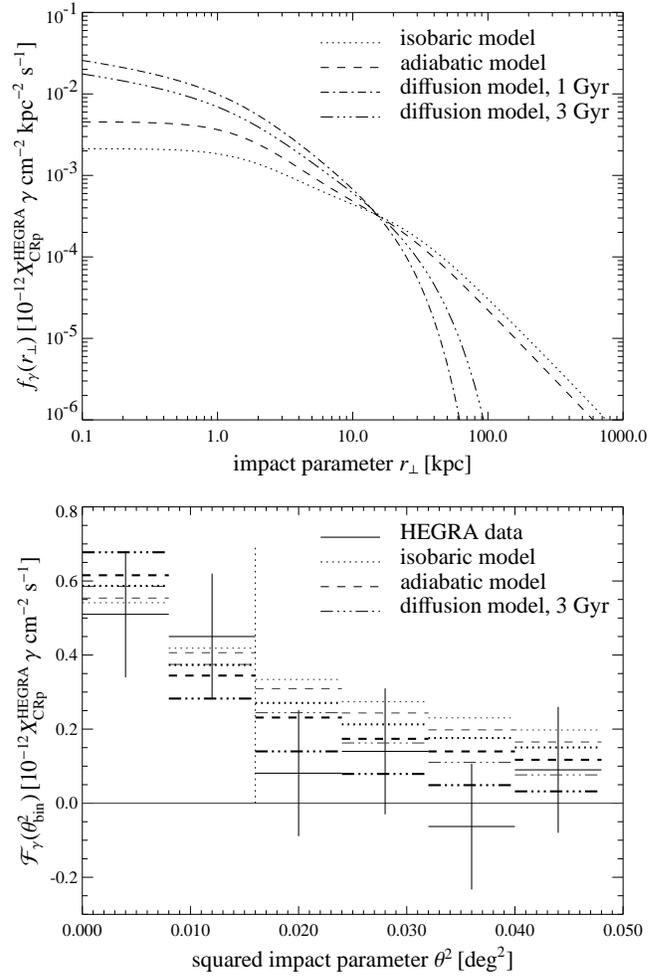}}
\caption{{\bf a)} Modeled $\gamma$-ray surface flux profiles $f_\gamma(r_\bot)$
  as function of impact parameter $r_\bot$ in our three different models for
  the spatial distribution of the CRp population.  They are normalized by
  comparing the integrated $\gamma$-ray flux above 730~GeV to HEGRA data of
  M~87 within the innermost two data points.  {\bf b)} Comparison of detected
  to integrated $\gamma$-ray flux $\mathcal{F}_\gamma(\theta^2)$ within the
  central aperture and the innermost annuli for different models of spatial CRp
  distribution as well as two different widths of the PSF. The thick black
  lines correspond to $\sigma = 0.05\degr$ whereas the thin grey lines are
  calculated for $\sigma = 0.08\degr$. The vertical dashed line separates the
  data from the noise level at a position corresponding to $r_\bot =
  37.5$~kpc.}
\label{fig: fig}
\end{figure}

The $\gamma$-ray flux profiles $f_\gamma(\vec{r}_\bot, E_\gamma >
E_\mathrm{thr})$ are obtained by integrating the $\gamma$-ray source function
$q_\gamma(\vec{r},E_\gamma)$ of eqn.~(\ref{q gamma}) above a threshold energy
$E_\mathrm{thr}$, successively projecting and convolving the spherically
symmetric profiles with the point spread function (PSF) of HEGRA,
$\mathrm{PSF}(\vec{r}_\bot) = \exp[-\vec{r}^2_\bot/(2\, \sigma^2)] /(2\, \pi\,
\sigma^2)$.  Resolution studies based on observations of the \object{Crab
  Nebula} with HEGRA indicate a width of $\sigma = 0.08\degr$, assuming a
differential spectral index of $\alpha = 2.7$ \citep{1997APh.....8....1D}.
However, for flatter power-law spectra being preferred by our hadronic
$\gamma$-ray model (see Sect.~\ref{sec:theory}), the width will be smaller
owing to increasing mean $\gamma$-ray energy. This leads to an increase of
$\gamma$-ray induced particles of the air shower and therefore better quality
of shower reconstruction according to a smaller relative Poissonian error.  For
a rough estimate, we rescaled the width of the PSF using the scaling of the
mean $\gamma$-ray energies above the instrumental threshold, yielding $\sigma =
0.05\degr$ with $\alpha = 2.2$.

The line-of-sight integration was performed out to a radius of $R_\mathrm{max}
\simeq 3$~Mpc which corresponds to the characteristic distance where the
$\beta$-model of electron densities is no longer applicable due to accretion
shocks of the cluster. The resulting $\gamma$-ray profiles are shown in
Fig.~\ref{fig: fig}.  The normalization of the surface fluxes depends on the
assumed scaling between CRp and thermal energy density, which is fixed by
comparing the integrated flux above 730~GeV to the innermost two $\gamma$-ray
flux data points of HEGRA \citep{2003A&A...403L...1A} corresponding to $\theta
= 0.126\degr$ or $r_\bot = 37.5$~kpc.  Although there are distinct
morphological differences visible in the three spatial CRp models, the
convolution with the PSF leads to very similar profiles for the expected HEGRA
$\gamma$-ray counts within the uncertainties. To demonstrate this, we compare
the integrated $\gamma$-ray flux $F_\gamma(E_\gamma > E_\mathrm{thr})$ for
different annuli of equal solid angle elements centered on the source to the
HEGRA data for the two different widths of the PSF discussed above
(Fig.~\ref{fig: fig}).

\section{Consequences for the CRp population in M~87}

By employing the technique described in Sect.~\ref{gamma profiles}, we explore
the consequences for the CRp scaling parameters $X_\crp$ and $\tilde{n}_{\crp,0}$ in the
particular models of CRp spatial distributions. The resulting values, shown in
Table~\ref{tab:Xcrp}, have been obtained using a PSF of width $\sigma =
0.05\degr$, however there are no significant changes in $X_\crp$ for $\sigma =
0.08\degr$.  The values of the CRp scaling parameter $X_\crp$ inferred from
M~87 are comparable to the one in our Galaxy, which is of order unity
\citep{1969SSRv....9..651P}.  Since the HEGRA $\gamma$-ray measurements probe
only the central region of the Virgo cluster which is dominated by the
elliptical radio galaxy M~87, a composition of ISM and ICM is observed,
potentially mixed by convective motion within the cooling flow
\citep{2001ApJ...554..261C}.  Therefore we expect $X_\crp$ to be smaller than
in our Galaxy, but significantly higher than upper limits obtained in nearby
cooling flow clusters, which are less than 20\% \citep{2003A&A...P}.

In the case of diffusion of CRp away from M~87, we are able to constrain the
averaged CRp luminosity $L_\crp$ of the central AGN by assuming a plausible
value for the diffusion coefficient $\kappa$.  The inferred values are of the
same order as instantaneous jet power estimates of M~87, $L_\mathrm{jet} \simeq
10^{43}\mbox{ erg}\mbox{ s}^{-1}$ \citep{1996ApJ...467..597B}. Thus, we limit a
combination of diffusion efficiency of CRp into the ambient thermal medium and
average jet power by this approach.

\begin{table}[t]
\caption[t]{
  Consequences for the CRp scaling parameter $X_\crp$ and $\tilde{n}_{\crp,0}$
  by comparing the integrated flux above 730 GeV to HEGRA data of the radio
  galaxy M~87 within the innermost two data points corresponding to $\theta =
  0.126\degr$. The spatial distribution of CRp is given by the isobaric,
  the adiabatic, and the diffusion model, respectively (see
  Sect.~\ref{CRpdistribution}). 
  In the first two cases the values are calculated for a CRp population with
  and without lower cutoff $p_\mathrm{min}$ while in the latter case two
  different lifetimes of the source have been considered. Note that the
  averaged CRp luminosity $L_\crp$ scales with $\kappa_{29} = \kappa/(10^{29}
  \mbox{ cm}^2 \mbox{ s}^{-1})$.}  
\vspace{-0.1 cm}
\begin{center}
\begin{tabular}{lcccc}
\hline \hline
\vphantom{\Large A}%
$\alpha_\mathrm{GeV}^\mathrm{TeV}$ 
& $X_\crp^\mathrm{isobaric}$ & $X_\crp^\mathrm{adiabatic}$
& $\tilde{n}_{\crp,0}~[\mbox{cm}^{-3}]$
& $L_\crp~\left[\kappa_{29}\mbox{ erg}/\mbox{s}\right]$ \\

\hline 
& \multicolumn{2}{l}{\underline{$p_\mathrm{min} = 0\mbox{ GeV} c^{-1}:$}}
& \multicolumn{2}{l}{\underline{$t = 1\mbox{ Gyr }\kappa_{29}^{-1}:$}}
\vphantom{\Large A} \\
\vphantom{\Large A}%
2.1  & 0.47 & 0.31 & $1.8 \times 10^{-7}$ & $1.1 \times 10^{43}$ \\
2.2  & 0.65 & 0.43 & $5.0 \times 10^{-7}$ & $1.6 \times 10^{43}$ \\
2.27 & 0.99 & 0.64 & $1.0 \times 10^{-6}$ & $2.4 \times 10^{43}$ \\
\hline 
& \multicolumn{2}{l}{\underline{$p_\mathrm{min} = 2\mbox{ GeV} c^{-1}:$}}
& \multicolumn{2}{l}{\underline{$t = 3\mbox{ Gyr }\kappa_{29}^{-1}:$}}
\vphantom{\Large A} \\
\vphantom{\Large A}%
2.1  & 0.42 & 0.28 & $1.2 \times 10^{-7}$ & $7.6 \times 10^{42}$ \\
2.2  & 0.52 & 0.34 & $3.4 \times 10^{-7}$ & $1.1 \times 10^{43}$ \\
2.27 & 0.73 & 0.47 & $6.8 \times 10^{-7}$ & $1.6 \times 10^{43}$ \\

\hline 
\end{tabular}
\end{center}
\label{tab:Xcrp}
\end{table}

Because of the scaling behavior of $X_\crp$ in the isobaric and adiabatic
models, we quantify the influence of a lower cutoff $p_\mathrm{min}$ on the
population of CRp due to Coulomb interactions in the plasma by taking the ratio
of CRp energy densities $\eps_\crp(p_\mathrm{min})$ with and without lower
cutoff (see eqn.~(\ref{energydensity})).  Such a cutoff yields lower values of
$X_\crp$ and therefore smaller contribution to the $\gamma$-ray flux in the
energy range of EGRET once the CRp momentum cutoff $p_\mathrm{min}$ exceeds the
kinematic threshold $p_\mathrm{thr} = 0.78\mbox{ GeV} c^{-1}$ of the hadronic
interaction. Thus, cooling of the CRp population allows for steeper power-law
distributions.

\section{Synchrotron emission by hadronic CRe}
Following the formalism described in \citet{2003A&A...P}, we compute the
synchrotron emission of CRe resulting from hadronic CRp interactions.
Integrating the expected radio surface brightness profiles over the solid angle
element corresponding to the $\gamma$-ray emission region and assuming magnetic
fields of the form $B(r) = 10\, \mu\mbox{G}\, [n_\mathrm{e}(r)
/n_\mathrm{e}(0)]^{0.5}$, we expect hadronic synchrotron fluxes $F_\nu = F_0\,
[\nu/(1.4 \mbox{ GHz})]^{-\alpha/2}$, where $F_0 = 11$~Jy and 16 Jy for $\alpha
= 2.1$ and 2.2. However, this hypothetical radio mini-halo is outshone by the
synchrotron emission of the relativistic jet, which shows a flux level of
$F_{1.4\,\mathrm{GHz}} = (220\pm 11)$~Jy \citep{1981A&AS...45..367K}.  The
hadronic radio surface profiles which are characterized by a smooth morphology
fall short by roughly one order of magnitude even at impact parameters of some
arc minutes compared to observed profiles of \citet{1996A&A...309L..19R}.

\section{Predictions for next generation \v{C}erenkov telescopes and GLAST}
There are three different scenarios predicting $\gamma$-ray emission from
objects like M~87, namely processed radiation of the relativistic outflow, dark
matter annihilation, and the hadronic scenario. The predictions of these types
of model differ predominantly in morphology, existing time-variability, and
spectral signatures. The radio galaxy M~87 which is well within the field of
view of the next generation \v{C}erenkov telescopes MAGIC\footnote{\tt
  http://hegra1.mppmu.mpg.de/MAGICWeb/}, HESS\footnote{\tt
  http://www.mpi-hd.mpg.de/hfm/HESS/HESS.html}, and VERITAS\footnote{\tt
  http://veritas.sao.arizona.edu/} should therefore serve as a unique source
for testing these scenarios. While the angular resolution of these telescopes
is comparable to the previously attained resolution, the flux sensitivities
have strongly improved. These developments should allow for $\gamma$-ray
spectroscopy by \v{C}erenkov experiments in the near future, providing the
opportunity of scrutinizing existing time-variation, and thus being able to
constrain different $\gamma$-ray emission scenarios.

The LAT instrument onboard GLAST\footnote{\tt
  http://glast.gsfc.nasa.gov/science/} will complement this research to even
lower energies ranging from 20~MeV up to 300~GeV.  Given a CRp population
described by a single power-law spectral index
$\alpha=\alpha_\mathrm{GeV}^\mathrm{TeV}$ extending from the GeV to TeV energy
regime as well as a CRp scaling parameter $X_\crp$ of Table~\ref{tab:Xcrp}, we
calculated the expected integrated $\gamma$-ray flux above 20~MeV.  In the
isobaric model, the $\gamma$-ray flux estimates are $\mathcal{F}_\gamma(>20
\mbox{ MeV}) /(\gamma \mbox{ cm}^{-2} \mbox{ s}^{-1})= 6.0\times 10^{-8},
1.3\times 10^{-7},\mbox{ and } 2.3 \times 10^{-7}$ for
$\alpha_\mathrm{GeV}^\mathrm{TeV} = 2.1,$ 2.2, and 2.27, respectively. This is
well above the sensitivity limit of GLAST.  The energy resolution of GLAST will
even provide the possibility to disentangle the pion decay induced signature
from inverse Compton emission of high-energetic electrons or positrons due to
the energy resolution which is better than 10\% and is sufficient to resolve
the pion decay induced peak in the $\gamma$-ray spectrum.

\section{Conclusion}
Using TeV $\gamma$-ray detections of M~87 by the HEGRA collaboration, it is for
the first time possible to constrain the CRp population of an elliptical
galaxy. By comparing to EGRET upper limits on the $\gamma$-ray emission, we
constrain the CRp spectral index $\alpha_\mathrm{GeV}^\mathrm{TeV}$, provided
the $\gamma$-ray emission is of hadronic origin and the population is described
by a single power-law ranging from the GeV to TeV energy regime.

By investigating three different models for the spatial distribution of the CRp
and applying those to realistic electron density and temperature profiles
obtained from X-ray observations, we calculate $\gamma$-ray flux profiles
resulting from hadronic CRp interactions with the thermal ambient gas using an
analytic formalism \citep{2003A&A...P}. After convolving with the HEGRA point
spread function, we compare the integrated $\gamma$-ray flux $F_\gamma(E_\gamma
> 730 \mbox{ GeV})$ for different annuli of equal solid angle elements centered
on the source. Based on the available data we find good morphological agreement
of all our models with these HEGRA \v{C}erenkov observations.

In the isobaric and adiabatic CRp model, the consequences for the CRp scaling
parameter $X_\crp$ drawn from normalization of our $\gamma$-ray flux profiles
to HEGRA observations yield slightly smaller values when comparing to our
Galaxy with $X_\crp \sim 1$, depending on the CRp spectral index. This is
because of the sensitivity of the observations to both the ISM of M~87 and the
ICM of the central cooling flow region of Virgo, where $X_\crp \lesssim 0.2$
\citep{2003A&A...P}.  Especially for $\alpha_\mathrm{GeV}^\mathrm{TeV} \simeq
2.1$ or lower momentum cutoffs of the CRp population due to Coulomb cooling
processes, we obtain smaller contributions of CRp pressure to the ambient
medium. By exploring our diffusion model and comparing our constraints on the
CRp luminosity $L_\crp$ to mechanical jet power estimates of M~87, we show the
ability of TeV $\gamma$-ray observations to constrain a combination of energy
fraction of CRp escaping from the radio plasma and average jet power of the
AGN.

The expected radio emission by hadronically produced CRe is roughly one order
of magnitude smaller compared to the synchrotron emission of the jet.
Therefore it will be a challenge for future radio observations to disentangle
the hadronic and jet emission components. Future \v{C}erenkov observations
should at least be able to severely constrain the parameter space of different
$\gamma$-ray emission scenarios. Finally, by investigating the $\gamma$-ray
flux in the energy regime of GLAST, we predict values which should allow to
scrutinize this hadronic model in contrast to other scenarios providing us with
the possibility of entering a new era of precision high energy cluster physics.

\begin{acknowledgements}
  We wish to thank Matthias Bartelmann, Sebastian Heinz, Francesco Miniati,
  Bj\"orn Malte Sch\"afer, Olaf Reimer, and an anonymous referee for carefully
  reading the manuscript and their constructive remarks.
\end{acknowledgements}

\bibliography{bibtex/chp}
\bibliographystyle{aa}

\end{document}